\documentclass[pra,showpacs,showkeys]{revtex4}
\usepackage{amsfonts,amsmath}
\usepackage[dvips]{graphics}
\begin{document}

\newcommand{\ket}[1]{\mathop{\left|#1\right>}\nolimits}       
\newcommand{\bra}[1]{\mathop{\left<#1\,\right|}\nolimits}       
\newcommand{\Tr}[1]{\mathop{{\mathrm{Tr}}_{#1}}}              
\newcommand{\sign}[1]{\mathop{{\mathrm{sign}}\left(#1\right)}} 
\newcommand{\braket}[2]{\langle #1 | #2 \rangle}
\newcommand{\ketbra}[2]{| #1\rangle\!\langle #2 |}
\newcommand{\nn}{\nonumber}
\newcommand{\vsp}{\vspace{.5cm}}
\newcommand{\p}{\prime}

\title{Optimal and covariant single-copy LOCC transformation between two two-qubit states}

\author{K. Br\'adler}
\affiliation{Instituto de F\'{\i}sica, Universidad Nacional
Aut\'onoma de M\'exico, Apdo. Postal 20-364, M\'exico 01000,
M\'exico}

\email{kbradler@epot.cz}

\date{\today}

\begin{abstract}
Given two two-qubit pure states characterized by their Schmidt
numbers we investigate an optimal strategy to convert the states
between themselves with respect to their local unitary invariance.
We discuss the efficiency of this transformation and its
connection to LOCC convertibility properties between two
single-copy quantum states. As an illustration of the investigated
transformations we present a communication protocol where in spite
of all expectations a shared maximally entangled pair between two
participants is the worst quantum resource.
\end{abstract}



\maketitle

\section{Introduction}

One of the greatest achievements of quantum information theory
(QIT) is the realization that quantum entanglement serves as a
resource for performing various communication tasks where Ekert's
scheme~\cite{ekert} for quantum key distribution (QKD) or quantum
teleportation~\cite{tele} are the most flagrant examples. Shortly
after, the question of equivalence of different multipartite
states came into question. Partially motivated by the security
issues in QKD (i.e. how to locally distill a shared non maximally
entangled or even noised quantum state to avoid any correlations
with a potential eavesdropper) the problem of LOCC~\cite{LOCC}
(local operation and classical communication)
convertibility~\cite{convertibility} became fundamental. We can
approach the question from two extremal sides. Namely, asking
whether two states are LOCC transformable in an asymptotic limit
or having just a single copy of an initial state at our disposal.
Both approaches brought the considerable progress in QIT. To name
just few, in the first regime, several measures of entanglement
were defined in terms of an asymptotic rate in which it is
possible to convert from/to a maximally entangled
state~\cite{ent_transf,LOCC}. In the second case, the connection
between the Schmidt number majorization~\cite{major} and LOCC
state transformation was discovered~\cite{nielsen_major} or new
classes of tripartite entangled states were presented~\cite{dur}.

We will treat with an interesting QIT paradigm which is so called
impossibility transformation (or `no-go~process'). There exist
several kinds of impossible transformations stratified by the fact
how the impossibility is fundamental. Quantum
cloning~\cite{cloning} or finding the orthogonal complement to a
given quantum state (universal NOT)~\cite{UNOT} belong to the
group of the highest stratum. This kind of impossibility comes
from basic principles of quantum mechanics~\cite{nocloning} and
can be performed just approximately~\cite{cloning,UNOT}. There are
also known other examples of fundamentally impossible
processes~\cite{other_impos}. In the lower level there exist
transformations which are not forbidden by the laws of quantum
mechanics but they are impossible under some artificially
augmented requirements. Typically, we consider only LOCC
operations as, for example, the above mentioned single-copy
transformation task~\cite{nielsen_major}. In this case, without
the LOCC constraint there is no problem to transform one pure
state to another without any limitations.

In this paper we use the methods of semidefinite
programming~\cite{semidef} to find an optimal and completely
positive (CP) map for LOCC single-copy pure state transformation
regarding its covariant properties. Covariance means that the
sought CP maps are universal in the sense that they do not change
their forms under the action of $SU(2)$ group (or their products)
on the input states. The covariance requirement was also added to
other quantum mechanical processes, compare e.g.~\cite{cov_other}.
In addition to the covariance, we require optimality meaning that
the output state produced by the CP LOCC map is maximally close to
the required target state. The closeness is measured by the value
of the fidelity between the actual output state and the desired
target state. As we will see, our problem of covariant and optimal
LOCC state transformation combines both kinds of the
impossibilities mentioned above.

The structure of the paper is the following. In
section~\ref{sec_intro} we recall some basics facts about the
isomorphism between quantum maps and the related group properties.
The main part of this paper can be found in
section~\ref{sec_transf} where the optimal LOCC single-copy state
transformation is investigated with the help of semidefinite
programming techniques. Section~\ref{sec_protocol} can be regarded
as an application of the studied problem where we present a
communication protocol for the LOCC transmission of a local
unitary operation from one branch of a shared two-qubit state to
the second one. We show that a maximally entangled pair does not
always need to be the best quantum communication resource. The
corresponding Kraus maps for the protocol are listed in Appendix.

\section{Methods}\label{sec_intro}

It is well known that there exists an isomorphism between
completely positive maps $\mathcal{M(\varrho)}$ and semidefinite
operators $R_\mathcal{M}$, first introduced by
Jamio{\l}kowski~\cite{jamiolk}
\begin{equation}\label{jamiolk}
     \mathcal{M}(\varrho_{in})
     =\Tr{in}\left[\left(\openone\otimes\varrho^T_{in}\right)R_\mathcal{M}\right]
     \Longleftrightarrow
     R_\mathcal{M}=\left(\mathcal{M}\otimes\openone\right)\left(P^{+}\right),
\end{equation}
where $P^{+}=\sum_{i,j=0}^{d-1}\ketbra{ii}{jj}$ is a maximally
entangled bipartite state of the dimension~$d^2$. The isomorphism
allows us to fulfill otherwise a difficult task of the
parametrization of all CP maps by putting a positivity condition
on the operator $R_\mathcal{M}$. Then, the parametrization problem
is computationally much more feasible. This is not the only
advantage the representation offers. As was shown
in~\cite{dariano+presti}, the representation is useful for the
description of quantum channels (so called covariant channels)
which we wish to optimize regarding some symmetry properties. More
precisely, having two representations $V_1,V_2$ of a unitary
group, the map $\mathcal{M(\varrho)}$ is said to be covariant if
$\mathcal{M(\varrho)}=V_2^\dagger\mathcal{M}(V_1\varrho
V_1^\dagger)V_2$. Inserting the covariance condition into
Eq.~(\ref{jamiolk}) and using the fact that the positive operator
$R_\mathcal{M}$ is unique, we get
\begin{equation}\label{Rcov}
    R_\mathcal{M}=(V_2^\dagger\otimes V_1^T)R_\mathcal{M}(V_2\otimes V_1^*)
    \Longleftrightarrow
    [R_\mathcal{M},V_2\otimes V_1^*]=0.
\end{equation}
The space occupied by $V_2\otimes V_1^*$ can be decomposed into a
direct sum of irreducible subspaces and from Schur's lemma follows
that $R_\mathcal{M}$ is a sum of the isomorphisms between all
equivalent irreducible representations. If we now consider the
fidelity equation in the Jamio{\l}kowski representation
\begin{equation}\label{fidel_cov_jamiolk}
    F=\Tr{}
    \left[
    \left(\varrho_{out}\otimes\varrho_{in}^T\right)R_\mathcal{M}
    \right]
\end{equation}
the task is reduced on finding the maximum of $F$ subject to
non-negativity of $R_\mathcal{M}$ and other constraints posed on
$R_\mathcal{M}$. In our case, it is the trace preserving condition
$\Tr{in}\left[R_\mathcal{M}\right]=\openone$ following
from~(\ref{jamiolk}). This can be easily reformulated as a
semidefinite program~\cite{semidef} and thus efficiently solved
using computers. Moreover, it is easy to put other conditions on
$R_\mathcal{M}$ such as partial positive transpose condition (PPT)
and they can be easily implemented as well~\cite{ppt}. Recall that
for two-qubit systems the PPT condition is equivalent to the LOCC
requirement. Note that the usefulness of the presented method was
already shown, for example, in connection with optimal and
covariant cloning~\cite{dobrr}.

In our calculation we employed the YALMIP
environment~\cite{YALMIP} equipped with the SeDuMi
solver~\cite{SeDuMi}. One of the advantages of semidefinite
programming is the indication of which parameters are zero. Then,
analytical solutions for the fidelity and even general forms of
the Kraus decomposition~\cite{kraus} of the CP map may be found.
In our problem, using the properties of the Jamio{\l}kowski
positive matrix $R_\mathcal{M}$ (which are stated as an almost
computer-ready theorem in~\cite{jami,zyck_book}) we derived the
corresponding Kraus operators as general as possible.

\section{Optimal and covariant single-copy LOCC state transformation}
\label{sec_transf}

Let us have an input and target state written in their Schmidt
forms $\ket{\chi}=a\ket{00}+\sqrt{1-a^2}\ket{11}$,
$\ket{\varphi}=c\ket{00}+\sqrt{1-c^2}\ket{11}$;
$a,c\in(0,1/\sqrt{2})$. It was shown~\cite{nielsen_major} that a
deterministic LOCC conversion $\ket{\chi}\to\ket{\varphi}$ is
possible iff $a\geq c$. If we want to go in the direction where
LOCC is not powerful enough we have basically two strategies at
our disposal. First, in some cases we may choose a probabilistic
strategy~\cite{concl_transf} also called conclusive conversion. As
an alternative, there exists a possible LOCC deterministic
transformation to a state which is in some sense closest to the
required one~\cite{faithful_transf}. Namely, it is a state in
which the fidelity with the target state is maximal. We will
follow a related way and find how an optimal covariant LOCC CP map
best approximates the ideal transformation
$\ket{\chi}\leftrightarrow\ket{\varphi}$. In the next sections, we
consider the following parameter space $a,c\in(0,1)$ both for
$\ket{\chi}$ and $\ket{\varphi}$.

Adopting the covariance considerations from the previous section
into our case we demand
\begin{equation}\label{fidel_cov}
F=\bra{\varphi}\mathcal{M(\ketbra{\chi}{\chi})}\ket{\varphi}
=\bra{\varphi^\p}\mathcal{M(\ketbra{\chi^\p}{\chi^\p})}\ket{\varphi^\p}
=F^\p,
\end{equation}
where
$V_1\ket{\chi}=\ket{\chi^\p},V_2\ket{\varphi}=\ket{\varphi^\p}$
and the covariance condition~(\ref{Rcov}) follows (note that quite
accidentally the condition is the same as in case of covariant
cloning).

\subsection{LOCC semicovariant transformation}

\begin{figure}[t]
\resizebox{14cm}{9cm}{\includegraphics{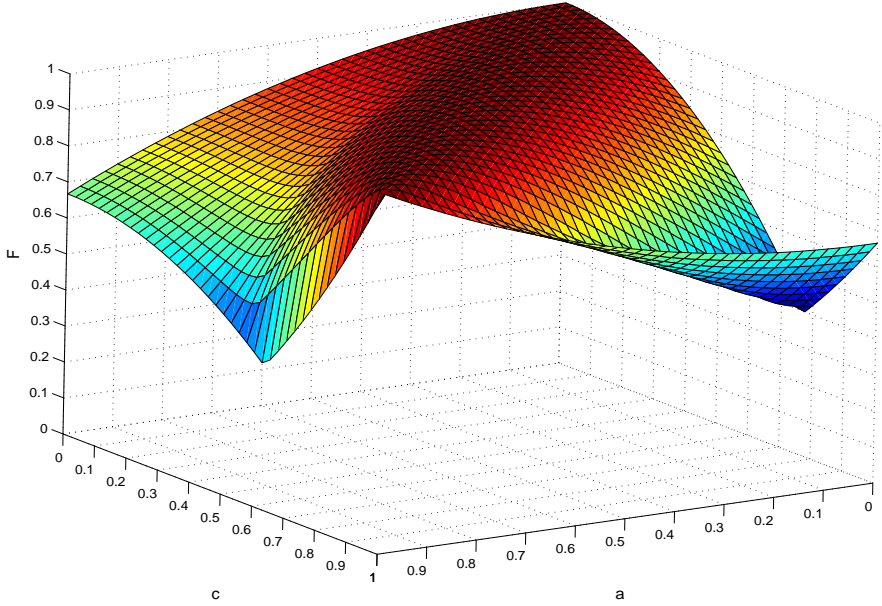}}
\caption{\label{fig_semi_fidel}The fidelity for the optimal and
locally semicovariant LOCC transformation between
$\ket{\chi}=a\ket{00}+\sqrt{1-a^2}\ket{11}$ and
$\ket{\varphi}=c\ket{00}+\sqrt{1-c^2}\ket{11}$.}
\end{figure}

\begin{figure}[h]
\resizebox{15cm}{6cm}{\includegraphics{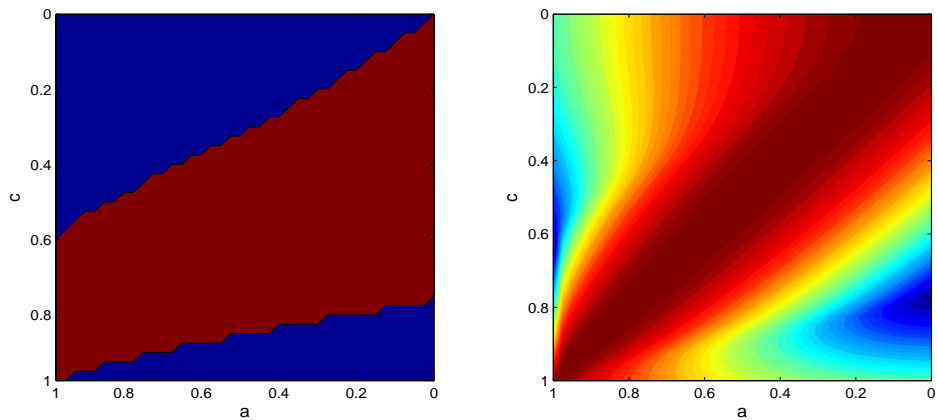}}
\caption{\label{fig_semi_fidel_identity}A 2D view on
Fig.~\ref{fig_semi_fidel} (on the right) together with the
indication where the trivial identity is the optimal map (the red
area on the left). The blue part corresponds to various non-unit
maps.}
\end{figure}

Firstly, we will be interested in how $\ket{\chi}$ can be
transformed if $V_1=V_2=\openone\otimes U$ where $U$ is a unitary
representation of $SU(2)$. In other words, we consider the
situation where the covariance is imposed on one branch of
$\ket{\chi}$ (we call it a {\it semicovariant} case). From
Eq.~(\ref{Rcov}) follows
\begin{equation}\label{semicov}
    [R_\mathcal{M},\openone\otimes U\otimes\openone\otimes U^*]=0
    \Longleftrightarrow
    [\tilde R_\mathcal{M},\openone\otimes\openone\otimes U\otimes U]=0,
\end{equation}
where $R_\mathcal{M}=S^\dagger\tilde R_\mathcal{M}S$ and
$S=\openone\otimes SW\!AP\otimes\sigma_Y$ where
$SW\!AP=\ketbra{00}{00}+\ketbra{01}{10}+\ketbra{10}{01}+\ketbra{11}{11}$
and $\sigma_Y$ is the Pauli $Y$ operator. With the unitarily
transformed rhs in Eq.~(\ref{semicov}) the decomposition is found
in a particularly simple way
\begin{equation}\label{semi_IR_decomp}
    \tilde R_\mathcal{M}=\bigoplus_{i,j=1}^4s_{ij}P_{S_{ij}}\oplus
    a_{ij}P_{A_{ij}},
\end{equation}
where $P_{S_{ij}},P_{A_{ij}}$ are isomorphisms between equivalent
symmetrical and antisymmetrical irreducible subspaces,
respectively. There are $32$ free complex parameters but we know
that $\tilde R_\mathcal{M}$ is a nonnegative operator. It follows
that $a_{ii},s_{ii}$ are real and
$a_{ij}=a_{ji}^*,s_{ij}=s_{ji}^*$. The number of free parameters
is thus reduced to $32$ real numbers. Maximizing the
fidelity~(\ref{fidel_cov_jamiolk}) for
$\varrho_{in}=\ketbra{\chi}{\chi},\varrho_{out}=\ketbra{\varphi}{\varphi}$
with this number of parameters is far from a possible analytical
solution but feasible in terms of semidefinite programming. For
$i\not=j$ it is advantageous to introduce the decomposition
$a_{ij}P_{A_{ij}}+a_{ij}^*P_{A_{ji}}=\Re[a_{ij}](P_{A_{ij}}+P_{A_{ji}})+\Im[a_{ij}](iP_{A_{ij}}-iP_{A_{ji}})$
and similarly for the symmetrical part. With the above defined
variables the fidelity to be maximized has the form (leaving out
the zero parameters)
\begin{equation}\label{fidelity_semi}
    F = {1\over2}\left(a^2c^2(s_{11}+a_{11})+(1-c^2)(1-a^2)(s_{44}+a_{44}))
      + c^2(1-a^2)s_{22}+(1-c^2)a^2s_{33}+ac\sqrt{(1-a^2)(1-c^2)}a^+_7\right),
\end{equation}
where $a^+_7=\Re[a_{41}]$. The resulting fidelity is depicted in
Fig.~\ref{fig_semi_fidel}.
First, we note that for $a\leq c$ the result corresponds to the
analytical result found in~\cite{faithful_transf} which, for our
bipartite case, has the form
\begin{equation}\label{semi_anal}
    F=\left(ac+\sqrt{(1-a^2)(1-c^2)}\right)^2.
\end{equation}
The reason is that the optimal fidelity found
in~\cite{faithful_transf} is dependent only on the Schmidt numbers
of the input and target state and thus it is automatically locally
covariant. If we do not consider the parameters of $R_\mathcal{M}$
which are shown to be zero (yielded from the semidefinite program)
a general form in the Kraus representation can be in principle
found ($R_\mathcal{M}$ can be diagonalized with the help of a
software for the symbolic manipulations). But it appears that this
decomposition is too complex and for our purpose it is not
necessary to present it. The only comment is deserved by the
identity map which covers the whole region of parameters where
Eq.~(\ref{semi_anal}) is valid. This is in contrast with the
original work~\cite{faithful_transf} where the map is not the
identity due to the knowledge of parameters $a,c$. In reality,
this trivial map appears to be the covariant and optimal map for a
bit larger region as depicted in
Fig.~\ref{fig_semi_fidel_identity}.
It follows that under the realm of the identity map no optimal
covariant CP map exists. The remaining part of the parameter space
of $a,c$ shows that in spite of the allowance of the perfect
deterministic conversion by the majorization criterion the
semicovariant transformation does not reach the maximal fidelity.
We intentionally left out the word LOCC because the second
interesting aspect is that for the whole parameter space the LOCC
condition is unnecessary. In other words, there are only LOCC
semicovariant transformations or the identity map which is also
(trivially) LOCC semicovariant. We confirm the existence of
another fundamental no-go process saying that it is not possible
to construct a CP map perfectly copying a partially or totally
unknown quantum state to a generally different quantum state even
if the majorization criterion allows us to do it (attention to the
related problem was called in~\cite{pati_imposs}). The
impossibility is easy to show by considering the following tiny
lemma valid not only for the investigated dimension $d=2$:\\
Let $M$ be a unitary and covariant map, i.e.
$\ket{\varphi}=M\ket{\chi}$ holds for two arbitrary qudits
$\ket{\chi},\ket{\varphi}$. Then, from the covariance follows
$MU\ket{\chi}=U\ket{\varphi}=UM\ket{\chi}\Longleftrightarrow[M,U]=0$.
We suppose that this holds for all $U\in SU(d)$ and then by one of
Schur's lemma $M=c\openone$. Considering the requirement of
unitarity of $M$ it follows $c=1$ and thus
$\ket{\varphi}=\ket{\chi}$.$\Box$\\
We confirmed this lemma in~Fig.~\ref{fig_semi_fidel} where the
fidelity is equal to one only if $a=c$ and we may reflect the
calculated optimal values of the fidelity as a refinement and
quantification how much is the above process impossible.

Note that the majorization criterion~\cite{nielsen_major} was
developed with respect to the degree of entanglement (the Schmidt
number) but relies on the complete knowledge of the converted
state what is at variance with the covariant requirement where no
particular state is preferred. The situation is a bit similar to
quantum cloning where if we know the preparation procedure of a
state to be cloned then there is no problem to make an arbitrary
number of its perfect copies.

Another worthy aspect is that the interval of $a$ and $c$ goes
from zero to one thus covering the target states with the same
Schmidt number more than once. Nevertheless, the fidelity is
different in such cases (compare e.g. the target states $\ket{00}$
and $\ket{11}$). In fact, to completely describe the
(semi)covariant properties of the type presented in this article
we should not distinguish input and target states by their Schmidt
numbers but rather to fully parametrize them in $SU(2)\otimes
SU(2)$ representation for every $a,c\in(0,1/\sqrt{2})$. But by
relying on the lemma above we expect that this situation does not
bring anything surprising into our discussion. Also, due to the
(semi)covariance we have actually described potentially
interesting transformations between
$\ket{\chi}=a\ket{01}+\sqrt{1-a^2}\ket{10}$ and
$\ket{\varphi}=c\ket{01}+\sqrt{1-c^2}\ket{10}$.

\subsection{Full LOCC covariant transformations}

As the second case we investigate a full local covariance where,
first, both qubits from an input two-qubit state $\ket{\chi}$  are
rotated simultaneously and, second, both qubits are rotated
independently. The covariance with respect to these two types of
rotation is required.

\begin{figure}[t]
\resizebox{14cm}{9cm}{\includegraphics{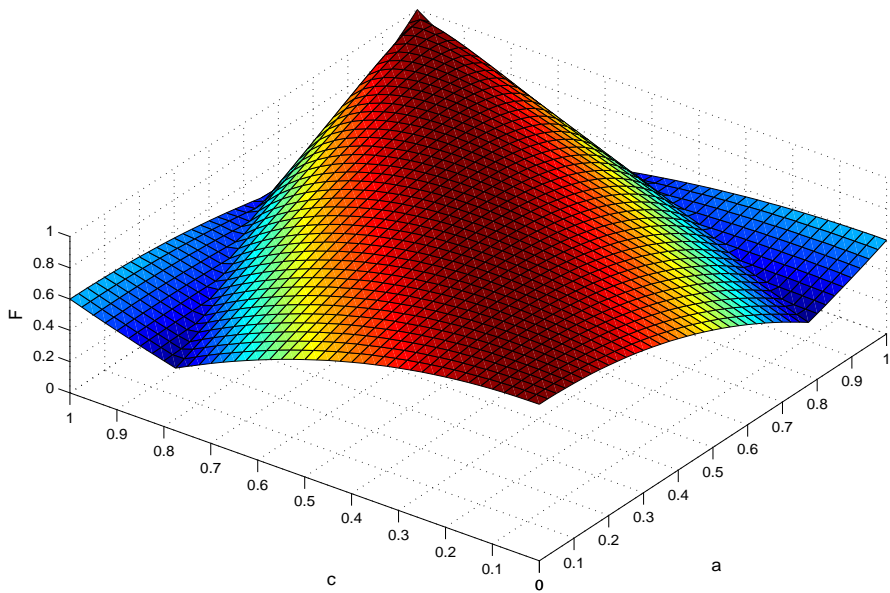}}
\caption{\label{fig_full_fidel}The fidelity for the optimal and
full locally covariant LOCC transformation between
$\ket{\chi}=a\ket{00}+\sqrt{1-a^2}\ket{11}$ and
$\ket{\varphi}=c\ket{00}+\sqrt{1-c^2}\ket{11}$.}
\end{figure}

The covariance condition in the first case is $V_1=V_2=U\otimes U$
and thus
\begin{equation}\label{fullcov}
    [R_\mathcal{M},U\otimes U\otimes U^*\otimes U^*]=0
    \Longleftrightarrow
    [\tilde R_\mathcal{M},U\otimes U\otimes U\otimes U]=0.
\end{equation}
Employing the fact that
\begin{equation}\label{tensor_dec}
    SU(2)^{\otimes4}_{j=1/2}=\bigoplus_{J=0}^2c_JD^{(J)}
\end{equation}
with $c_J\in(2,3,1)$ we find the basis vectors of all irreducible
subspaces (summarized in Tab.~\ref{tab_irreps}) and construct
isomorphisms $P$ between equivalent species
\begin{equation}\label{full_IR_decomp}
    \tilde
    R_\mathcal{M}=\bigoplus_{J=0}^2\bigoplus_{k,l=1}^{c_J}d_{Jkl}P_{D^{(J)}_{kl}}.
\end{equation}
Choosing the parameters $d_{Jkl}$ we require $R_\mathcal{M}$ to be
a semidefinite matrix. We calculate the fidelity for the same kind
of input/target states from the previous subsection yielding
\begin{equation}\label{fidelity_full}
F =
\left(ac+\sqrt{(1-a^2)(1-c^2)}\right)^2\left({1\over3}d_{022}+{1\over6}d_{211}\right)
+ \left(c^2(1-a^2)+(1-c^2)a^2\right)d_{211}.
\end{equation}
Running an appropriate semidefinite program for maximizing $F$ we
are able to get analytical results both for the fidelity and the
CP map in the Kraus form. It appears that many of the coefficients
$d_{Jkl}$ are zero and thus Eq.~(\ref{fidelity_full}) simplifies
as well as the constraints given by the trace preserving
condition. As far as the LOCC condition the situation here is that
the CP maps with and without the posed condition are different but
both give the same optimal fidelity. It can be shown that the LOCC
condition in this case is just a dummy constraint determining the
value of a free parameter in the resulting map (see the parameter
$d_{011}$ in Eq.~(\ref{full_kraus})). Then
\begin{equation}\label{fidelity_full_anal}
F = \max{\left[\left(ac+\sqrt{(1-a^2)(1-c^2)}\right)^2,
{1\over10}\left(ac+\sqrt{(1-a^2)(1-c^2)}\right)^2+{3\over5}\left(c^2(1-a^2)+a^2(1-c^2)\right)\right]}
\end{equation}
and the corresponding graph is in Fig.~\ref{fig_full_fidel}.
It is noteworthy that there are just two types of CP~covariant
maps for two investigated intervals of $a,c$ corresponding to the
different fidelity functions in~(\ref{fidelity_full_anal}). The
identity map is the first one and the conclusion from the previous
case holds. The second map is described by the set of the Kraus
operators
\begin{eqnarray}\label{full_kraus}
&   A_1=\sqrt{1-d_{011}\over3}\begin{pmatrix}
      0 & -1 & 1 & 0 \\
      0 & 0 & 0 & 0 \\
      0 & 0 & 0 & 0 \\
      0 & 0 & 0 & 0 \\
    \end{pmatrix},
     A_2=\sqrt{1-d_{011}\over3}\begin{pmatrix}
      0 & 0 & 0 & 0 \\
      0 & 0 & 0 & 0 \\
      0 & 0 & 0 & 0 \\
      0 & 1 & -1 & 0 \\
    \end{pmatrix},
     A_3=\sqrt{1-d_{011}\over12}\begin{pmatrix}
      0 & 0 & 0 & 0 \\
      0 & -1 & 1 & 0 \\
      0 & -1 & 1 & 0 \\
      0 & 0 & 0 & 0 \\
    \end{pmatrix},\nonumber\\
    &A_4={\sqrt{d_{011}}\over2}\begin{pmatrix}
      0 & 0 & 0 & 0 \\
      0 & 1 & -1 & 0 \\
      0 & -1 & 1 & 0 \\
      0 & 0 & 0 & 0 \\
    \end{pmatrix},
    A_5={1\over\sqrt{10}}\begin{pmatrix}
      1 & 0 & 0 & 0 \\
      0 & -1 & -1 & 0 \\
      0 & -1 & -1 & 0 \\
      0 & 0 & 0 & 1 \\
    \end{pmatrix},
    A_6=\sqrt{3\over20}\begin{pmatrix}
      0 & 0 & 0 & 0 \\
      -1 & 0 & 0 & 0 \\
      -1 & 0 & 0 & 0 \\
      0 & 1 & 1 & 0 \\
    \end{pmatrix},\nonumber\\
    &A_7=\sqrt{3\over20}\begin{pmatrix}
      0 & -1 & -1 & 0 \\
      0 & 0 & 0 & 1 \\
      0 & 0 & 0 & 1 \\
      0 & 0 & 0 & 0 \\
    \end{pmatrix},
    A_8=\sqrt{3\over5}\begin{pmatrix}
      0 & 0 & 0 & 1 \\
      0 & 0 & 0 & 0 \\
      0 & 0 & 0 & 0 \\
      0 & 0 & 0 & 0 \\
    \end{pmatrix},
    A_9=\sqrt{3\over5}\begin{pmatrix}
      0 & 0 & 0 & 0 \\
      0 & 0 & 0 & 0 \\
      0 & 0 & 0 & 0 \\
      1 & 0 & 0 & 0 \\
    \end{pmatrix},
\end{eqnarray}
where $d_{011}$ is a free parameter from the
decomposition~(\ref{full_IR_decomp}). The trace-preserving
condition $\sum_{i=1}^9A_i^\dagger A_i=\openone$ is
satisfied~\footnote{Decomposition~(\ref{full_kraus}) is not in a
visible LOCC form but we know that Kraus maps are not
unique~\cite{kraus} as well as the corresponding positive matrices
in the Jami\l kowski representation. However, due to the PPT
condition laid on $R_\mathcal{M}$ the particular PPT (for
two-qubit states thus LOCC) Kraus decomposition can be derived.}.
\begin{table}
\caption{\label{tab_irreps}Orthogonal basis vectors of all
irreducible subspaces of $SU(2)^{\otimes4}_{j=1/2}$.}
\begin{ruledtabular}
\begin{tabular}{ccc}
  Total momentum $J$ & Irreducible subspace $D^{(J)}_{kl}$ & Basis vectors \\
  \hline
  0 & $D^{(0)}_{11}$ &  ${1\over2}\ket{01-10}\ket{01-10}$\\
  \hline
  0 & $D^{(0)}_{22}$ &  ${1\over\sqrt{3}}\left(\ket{0011}-{1\over2}\ket{01+10}\ket{01+10}+\ket{1100}\right)$\\
  \hline
  1 & $D^{(1)}_{11}$ &  ${1\over\sqrt{2}}\ket{01-10}\ket{00}$\\
    &                &  ${1\over2}\ket{01-10}\ket{01+10}$\\
    &                &  ${1\over\sqrt{2}}\ket{01-10}\ket{11}$\\
  \hline
  1 & $D^{(1)}_{22}$ &  ${1\over\sqrt{2}}\ket{00}\ket{01-10}$\\
    &                &  ${1\over2}\ket{01+10}\ket{01-10}$\\
    &                &  ${1\over\sqrt{2}}\ket{11}\ket{01-10}$\\
  \hline
  1 & $D^{(1)}_{33}$ &  $-{1\over2}\left(\ket{00}\ket{01+10}-\ket{01+10}\ket{00}\right)$\\
    &                &  $-{1\over\sqrt{2}}\left(\ket{0011}-\ket{1100}\right)$\\
    &                &  $-{1\over2}\left(\ket{01+10}\ket{11}-\ket{11}\ket{01+10}\right)$\\
  \hline
  2 & $D^{(2)}_{11}$ &  $\ket{0000}$\\
    &                &  ${1\over2}\left(\ket{00}\ket{01+10}+\ket{01+10}\ket{00}\right)$\\
    &                &  ${1\over\sqrt{6}}\left(\ket{0011}+\ket{1100}+\ket{01+10}\ket{01+10}\right)$\\
    &                &  ${1\over2}\left(\ket{01+10}\ket{11}+\ket{11}\ket{01+10}\right)$\\
    &                &  $\ket{1111}$\\
\end{tabular}
\end{ruledtabular}
\end{table}

Let us proceed to the second case where we consider independent
unitary rotations on both qubits of the pair, that is
$V_1=V_2=U_1\otimes U_2$. Derived analogously as before, it
follows
\begin{equation}\label{pathocov}
    [\tilde R_\mathcal{M},U_1\otimes U_1\otimes U_2\otimes U_2]=0
\end{equation}
with the decomposition in a particularly simple form
\begin{equation}\label{patho_IR_decomp}
    \tilde
    R_\mathcal{M}=p_1P_A\otimes P_A+p_2P_A\otimes P_S+p_3P_S\otimes P_A+p_4P_S\otimes P_S,
\end{equation}
where $P_A,P_S$ are the projectors into asymmetrical and
symmetrical subspaces, respectively~\cite{RDD_comment}. The
resulting fidelity equation (again independent on the LOCC
condition) can be derived analytically
\begin{equation}\label{fidelity_full_anal2}
    F=\max{\left[\left(ac+\sqrt{(1-a^2)(1-c^2)}\right)^2,
      {1\over9}\left(ac+\sqrt{(1-a^2)(1-c^2)}\right)^2+{4\over9}\left(c^2(1-a^2)+a^2(1-c^2)\right)\right]}
\end{equation}
with the picture looking similarly as
in~Fig.~\ref{fig_full_fidel}. The achieved fidelity is even lower
due to the stronger requirements on the covariance properties in
Eq.~(\ref{pathocov}) in comparison with Eq.~(\ref{fullcov}). As in
the previous case, there are two maps  for two different fidelity
functions, one of them being the identity map. The Kraus
decomposition of the nontrivial map is
\begin{equation}\label{full_kraus_2}
    A_1={1\over3}\begin{pmatrix}
      1 & 0 & 0 & 0 \\
      0 & -1 & 0 & 0 \\
      0 & 0 & -1 & 0 \\
      0 & 0 & 0 & 1 \\
    \end{pmatrix},
    A_2=\begin{pmatrix}
      0 & 0 & 0 & 0 \\
      -{\sqrt{2}\over3} & 0 & 0 & 0 \\
      0 & 0 & 0 & 0 \\
      0 & 0 & {\sqrt{2}\over3} & 0 \\
    \end{pmatrix},
    A_3=\begin{pmatrix}
      0 & 0 & 0 & 0 \\
      0 & 0 & 0 & 0 \\
      -{\sqrt{2}\over3} & 0 & 0 & 0 \\
      0 & {\sqrt{2}\over3} & 0 & 0 \\
    \end{pmatrix},
    A_4=\begin{pmatrix}
      0 & 0 & 0 & 0 \\
      0 & 0 & 0 & 0 \\
      0 & 0 & 0 & 0 \\
      {2\over3} & 0 & 0 & 0 \\
    \end{pmatrix},
    A_5=\begin{pmatrix}
      0 & 0 & 0 & 0 \\
      0 & 0 & 0 & 0 \\
      0 & {2\over3} & 0 & 0 \\
      0 & 0 & 0 & 0 \\
    \end{pmatrix}
\end{equation}
and
$A_6=A_2^\dagger,A_7=A_3^\dagger,A_8=A_5^\dagger,A_9=A_4^\dagger$.

\section{Covariant LOCC communication protocol}\label{sec_protocol}

\begin{figure}[t]
\resizebox{10cm}{6.5cm}{\includegraphics{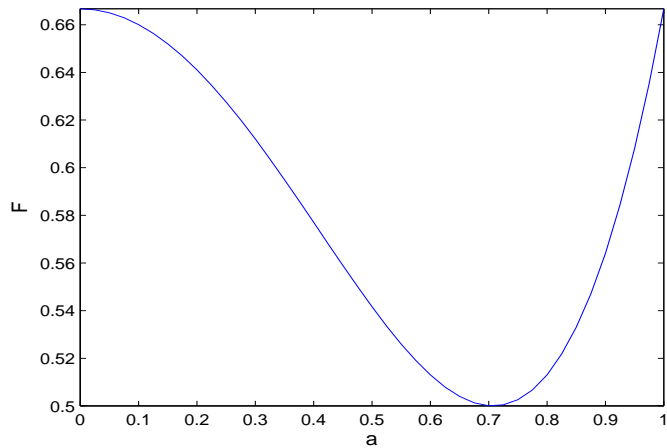}}
\caption{\label{fig_protocol_fidel}The fidelity of the protocol
for `handing over' a local unitary operation between branches of a
partially entangled two-qubit pair. The entanglement of the shared
pair is characterized by the Schmidt number $a$.}
\end{figure}

Let us try to apply the previous considerations to the solution of
the following communication problem. Suppose that the two-qubit
state $\ket{\chi}=a\ket{00}+\sqrt{1-a^2}\ket{11}$ was locally and
unitarily modified on Alice's side and then distributed between
Alice and Bob. Next imagine that the distributor of this state is
confused and oblivious and he wanted originally to modify Bob's
part of the state. Moreover, he forgot which unitary modification
was done. Since Alice and Bob are separated the only possibility
to rectify the distributor's mistake is LOCC communication between
them. In other words, they would like to perform the following
transformation
\begin{equation}\label{protocol}
\ket{\chi^\p}=(U\otimes\openone)\ket{\chi}
\stackrel{LOCC}{\to}(\openone\otimes U)\ket{\chi}=\ket{\varphi^\p}
\end{equation}
such that the LOCC transformation will be equally and maximally
successful irrespective of $U$. Generally, this is the problem of
sending an unknown local unitary operation between branches of a
shared bipartite state. Notice that if $\ket{\chi}$ is a maximally
entangled state then the task changes to finding a transposition
of the unitary operation $U$ due to the well known relation
\begin{equation}\label{known_relation}
(U\otimes\openone)\ket{00+11}=(\openone\otimes U^T)\ket{00+11}.
\end{equation}
The covariant condition in the Jamio{\l}kowski representation
reads
\begin{equation}\label{protcov}
   [R_\mathcal{M},\openone\otimes U\otimes U^*\otimes\openone]=0
\end{equation}
using decomposition~(\ref{semi_IR_decomp}) and the unitary
modification $\tilde R_\mathcal{M}=SR_\mathcal{M}S^\dagger$ with
$S=(\openone\otimes
SW\!AP\otimes\sigma_Y)(\openone\otimes\openone\otimes SW\!AP)$.
Again, the figure of merit is the fidelity which now has the form
\begin{equation}\label{fidelity_protocol}
    F={1\over2}\left(a^4(s_{11}+a_{11})+(1-a^2)^2(s_{44}+a_{44})\right)
    +a^2(1-a^2)(s_{22}+s_{33}+a_7^+-s_7^+),
\end{equation}
where $a^+_7=\Re[a_{41}],s^+_7=\Re[s_{41}]$. One may find a
general form of this map in terms of the Kraus operators in
Appendix. If we first run the corresponding semidefinite program
without the LOCC condition we get the fidelity equal to one for
all $a$. This has a reasonable explanation because if we allow the
nonlocal operations there exists a universal and always successful
unitary operation -- SW\!AP. The inspection of the particular
$R_\mathcal{M}$ confirms this inference. After imposing the LOCC
condition the resulting fidelity is depicted in
Fig~\ref{fig_protocol_fidel}.
This result is noteworthy because we see that the LOCC CP map is
the most successful for the factorized states ($a=0,1\sim F=2/3$)
while it holds $F=1/2$ for the maximally entangled states. The
reason lies in Eqs.~(\ref{known_relation}) and~(\ref{protocol}).
If $\ket{\chi}$ is a maximally entangled state then a local
unitary action passes the whole local orbit whereas for
non-maximally entangled states the unitary action on one branch is
not sufficient for the attainment of all possible partially
entangled states characterized by the same Schmidt number $a$. We
may conclude with an intriguing claim that in case of our protocol
it is better for Alice and Bob to share a factorized state instead
of a maximally entangled state. Let us stress that the optimal map
is not trivially identical for any value of the parameter $a$ in
the input state $\ket{\chi}$.

\section{Conclusion}

In this work we studied the LOCC transformations between two-qubit
bipartite states characterized by their Schmidt numbers. In
addition to the obvious CP requirement, we looked for the
covariant maps which maximize the fidelity between an input and a
target state. Moreover, we supposed that we had just a single copy
of the input state at our disposal. The studied covariance can be
divided into two groups: so called semicovariance where we
required the independence of the input state regarding the action
of $SU(2)$ representation on one of the input qubits. The second
investigated possibility were two cases of full covariance
condition where the independence and optimality of the state
transformation had been examined with respect to two (equivalent
and nonequivalent) $SU(2)$ representations acting on both branches
of the input bipartite state.

We employed the methods of semidefinite programming which, in
spite of being a numerical method, enables us to find totally or
partially general analytical solutions for the fidelity and for
the corresponding LOCC CP maps. We have found that, first, due to
the covariance conditions there are no possible perfect state
transformations even if the majorization criterion allows them and
with the calculated optimal fidelity we quantified the `maximal
allowance' of the considered transformations. Second, we have
shown that there only exist LOCC covariant transformations. Hence,
since this condition is unnecessary this kind of transformation
can be rated as another basic process forbidden by the laws of
quantum mechanics. We have also connected our work with the
earlier works on so called faithful single-copy state
transformations~\cite{faithful_transf}. Notably, for the
corresponding subset of the investigated parameter area the same
analytical results for the fidelity were derived but under the
local unitary covariant circumstances. Consequently, the forms of
the particular CP maps are different from previously derived
putting this problem into a different perspective.

Finally, we illustrated these methods on an application of the
communication protocol for LOCC `handing over' of a local unitary
operation from one branch of a shared two-qubit bipartite state to
another without its actual knowledge. Intriguingly, is has been
shown that the best results (in terms of the fidelity between an
input and a target state) are achieved if both parties share one
of the considered factorized states $\ket{00}$ or $\ket{11}$ and
not the maximally entangled state.

Even if for general multipartite states the PPT condition used
here is not equivalent to the LOCC condition, the described
methods might be useful for this kind of study as well, for
example, to help clarifying the role of the PPT operations and the
transformation properties of these states.

\begin{acknowledgments}
The author is very indebted for discussions and support from
R.~J\'auregui and for comments from R. Demkowicz-Dobrza\'nski.
\end{acknowledgments}

\appendix
\section{}

Considering
\begin{eqnarray}\label{app_peqs}
    &p_{1,2}={\frac {-\,s_{11}+\,s_{44}\pm\,\sqrt
    {s_{11}^{2}-2\,s_{11}s_{44}+s_{44}^{2}+4\,(s_7^+)^{2}}}{2s_7^+}}\\
    &p_{3,4}={\frac {-\,a_{11}+\,a_{44}\pm\,\sqrt
    {a_{11}^{2}-2\,a_{11}a_{44}+a_{44}^{2}+4\,(a_7^+)^{2}}}{2a_7^+}}
\end{eqnarray}
and
\begin{eqnarray}
  d_1 &=& {1\over\sqrt{2}}\left({s_{11}+\,s_{44}+\,\sqrt{s_{11}^{2}-2\,s_{11}s_{44}+s_{44}^{2}+4\,(s_7^+)^{2}}}\right)^{1/2} \\
  d_2 &=& {1\over\sqrt{2}}\left({s_{11}+\,s_{44}-\,\sqrt{s_{11}^{2}-2\,s_{11}s_{44}+s_{44}^{2}+4\,(s_7^+)^{2}}}\right)^{1/2} \\
  d_3 &=& {1\over\sqrt{2}}\left({a_{11}+\,a_{44}+\,\sqrt{a_{11}^{2}-2\,a_{11}a_{44}+a_{44}^{2}+4\,(a_7^+)^{2}}}\right)^{1/2} \\
  d_4 &=& {1\over\sqrt{2}}\left({a_{11}+\,a_{44}-\,\sqrt{a_{11}^{2}-2\,a_{11}a_{44}+a_{44}^{2}+4\,(a_7^+)^{2}}}\right)^{1/2} \\
  d_5 &=& \sqrt{s_{22}} \\
  d_6 &=& \sqrt{s_{33}}
\end{eqnarray}
we may write the Kraus operators for the problem in
Sec.~\ref{sec_protocol} as
\begin{eqnarray}\label{mega_kraus}
&   A_1={d_1\over\sqrt{2}}{1\over\sqrt{1+p_1^2}}\begin{pmatrix}
      {p_1} & 0 & 0 & 0 \\
      0 & 0 & -{p_1} & 0 \\
      0 & -1 & 0 & 0 \\
      0 & 0 & 0 & 1 \\
    \end{pmatrix},
     A_2={d_1\over\sqrt{2}}{1\over\sqrt{1+p_1^2}}\begin{pmatrix}
      0 & 0 & 0 & 0 \\
      -1 & 0 & 0 & 0 \\
      0 & 0 & 0 & 0 \\
      0 & 1 & 0 & 0 \\
    \end{pmatrix},\nonumber\\
     &A_4={d_2\over\sqrt{2}}{\sign{p_2-p_1}\over\sqrt{1+p_1^2}}\begin{pmatrix}
      1 & 0 & 0 & 0 \\
      0 & 0 & -1 & 0 \\
      0 & {p_1} & 0 & 0 \\
      0 & 0 & 0 & -{p_1} \\
    \end{pmatrix},
    A_5={d_2\over\sqrt{2}}{\sign{p_2-p_1}\over\sqrt{1+p_1^2}}\begin{pmatrix}
      0 & 0 & 0 & 0 \\
      -1 & 0 & 0 & 0 \\
      0 & 0 & 0 & 0 \\
      0 & -1 & 0 & 0 \\
    \end{pmatrix},\nonumber\\
    &A_7={d_3\over\sqrt{2}}{1\over\sqrt{1+p_3^2}}\begin{pmatrix}
      {p_3} & 0 & 0 & 0 \\
      0 & 0 & {p_3} & 0 \\
      0 & 1 & 0 & 0 \\
      0 & 0 & 0 & 1 \\
    \end{pmatrix},
    A_8={d_4\over\sqrt{2}}{\sign{p_3-p_4}\over\sqrt{1+p_3^2}}\begin{pmatrix}
      -1 & 0 & 0 & 0 \\
      0 & 0 & -1 & 0 \\
      0 & {p_3} & 0 & 0 \\
      0 & 0 & 0 & {p_3} \\
    \end{pmatrix},\nonumber\\
    &A_9={d_5\over\sqrt{2}}\begin{pmatrix}
      0 & -1 & 0 & 0 \\
      0 & 0 & 0 & 1 \\
      0 & 0 & 0 & 0 \\
      0 & 0 & 0 & 0 \\
    \end{pmatrix},
    A_{10}=d_5\begin{pmatrix}
      0 & 0 & 0 & 0 \\
      0 & 1 & 0 & 0 \\
      0 & 0 & 0 & 0 \\
      0 & 0 & 0 & 0 \\
    \end{pmatrix},
    A_{11}=d_5\begin{pmatrix}
      0 & 0 & 0 & 1 \\
      0 & 0 & 0 & 0 \\
      0 & 0 & 0 & 0 \\
      0 & 0 & 0 & 0 \\
    \end{pmatrix},\nonumber\\
     & A_{12}={d_6\over\sqrt{2}}\begin{pmatrix}
      0 & 0 & 0 & 0 \\
      -1 & 0 & 0 & 0 \\
      0 & 0 & 0 & 0 \\
      0 & 1 & 0 & 0 \\
         \end{pmatrix},
    A_{13}=d_6\begin{pmatrix}
      0 & 0 & 0 & 0 \\
      0 & 0 & 0 & 0 \\
      0 & 0 & 0 & 0 \\
      1 & 0 & 0 & 0 \\
         \end{pmatrix},
    A_{14}=d_6\begin{pmatrix}
      0 & 0 & 0 & 0 \\
      0 & 0 & 0 & 0 \\
      0 & 0 & 1 & 0 \\
      0 & 0 & 0 & 0 \\
\end{pmatrix},
\end{eqnarray}
and $A_3=A_2^\dagger,A_6=A_5^\dagger$. The maps satisfy
$\sum_{i=1}^{14}A_i^\dagger A_i=\openone$ if the trace preserving
condition on the Jamio{\l}kowski map is posed. Similarly to
Eq.~(\ref{full_kraus}), the Kraus operators are not in their
apparent LOCC form but can be transformed into it.

\end{document}